\documentclass[useAMS,usenatbib]{mn2e}
\usepackage{graphics,epsfig,xtab,lscape,txfonts,color}
\usepackage[english]{babel}

\newcommand{\mr}{\mathrm}

\def\eg{e.\,g.}                                      

\def\xmm{\textit{XMM-Newton}}
\def\swift{\textit{Swift}}

\def\gx339{GX\,339-4}
\def\h1743{H\,1743-322}

\def\gs{GS\,1354--64}

\title[2015 outburst of GS1354--64]{The 2015 hard-state only outburst of \gs\thanks{Based on observations obtained with \xmm, an ESA science mission with instruments and contributions directly funded by ESA Member States and NASA.}}
\author[H. Stiele, A. Kong]{H.\ Stiele$^{1}$\thanks{E-mail:
hstiele@mx.nthu.edu.tw}, A.\ K.\ H.\ Kong$^{1}$ \\
$^{1}$National Tsing Hua University, Department of Physics and Institute of Astronomy, No.~101 Sect.~2 Kuang-Fu Road,  30013, Hsinchu, Taiwan}
\begin{document}

\date{Accepted 2016 April 13. Received 2016 April 13; in original form 2016 February 3}

\pagerange{\pageref{firstpage}--\pageref{lastpage}} \pubyear{2016}

\maketitle

\label{firstpage}

\begin{abstract}
Since its outburst in 1997 \gs\ stayed in quiescence. In June 2015 renewed activity of \gs\ was observed. Based on our analysis of energy spectra and timing properties obtained from \swift/XRT monitoring data we found that \gs\ stayed in the hard state during the entire outburst. Such a hard state only (or ``failed'' outburst) has also been observed in 1997.
In addition, we analysed an \xmm\ observation taken on August 6th. We compared variability on long and short time scales using covariance ratio and found that the ratio showed a decrease towards lower energies instead of the increase that has been found in other black hole X-ray binaries. There are now two sources (\h1743\ and \gs) that do not show an increase towards lower energies in their covariance ratio. Both sources have been observed during ``failed'' outbursts and showed photon indices much harder than what is usually observed in black hole X-ray binaries. 
\end{abstract}


\section{Introduction}

\gs, located at a distance of 25 -- 61 kpc \citep{2009ApJS..181..238C}, was discovered by the all sky monitor onboard the \textit{Ginga} satellite \citep{1987Natur.326..322S} during an outburst in 1987 \citep{1987IAUC.4342....1M}. During this outburst \gs\ showed spectra dominated by thermal emission form an accretion disc with a power law tail at energies above 10 keV. This kind of spectra is typically observed during the high-soft state (HSS) of a black hole X-ray binary (BHB). There are two other transient sources -- Cen X-2 \citep{1971NPhS..229..229F} and MX 1353-64 \citep{1979ApJS...39..573M} -- which have been discovered earlier and which positions are consistent with the one of \gs. Cen X-2 is one of the brightest X-ray transients and was also observed during HSS \citep{1995xrbi.nasa..126T}, while MX 1353-64 was observed during low-hard state (LHS) when the spectrum is dominated by non-thermal emission of a hot Comptonizing plasma. In case these three sources are the same source, it would show different spectral states during an outburst like most known BHBs do. Another outburst of \gs\ has been detected by RXTE in 1997. During this outburst \gs\ was about a factor 3 fainter than during the 1987 outburst  \citep{2000ApJ...530..955R}. Analysis of RXTE spectra and power density spectra showed that \gs\ stayed in the LHS during the entire outburst \citep{2000ApJ...530..955R,2001MNRAS.323..517B}. 

\gs\ showed another outburst between June and September 2015 that was followed by \swift/XRT observations \citep{2015ATel.7612....1M}. During its 2015 outburst \gs\ reached about the same brightness as during the 1997 outburst. The spectral and timing analysis presented in this paper showed that \gs\ stayed in the LHS during the entire outburst.

\section[]{Observations and data analysis}
\label{Sec:obs}
In this paper, we present a comprehensive study of the spectral and temporal variability properties of \gs\ observed during its 2015 outburst. Renewed activity of the source was detected in the \swift/BAT and MAXI monitoring observations. 

\subsection{\swift}
A series of \swift/XRT observations have been taken to trace the outburst evolution. Here we analysed all \swift/XRT observations of \gs\ taken in windowed timing mode between June 10th and September 20th. We extracted energy spectra of each observation using the online data analysis tools provided by the Leicester \swift\ data centre\footnote{http://www.swift.ac.uk/user\_objects/}, including single pixel events only. In addition, we extracted power density spectra (PDS) in the 0.3 -- 10 keV energy band, following the procedure outlined in \citet{2006MNRAS.367.1113B}. We subtracted the contribution due to Poissonian noise \citep{1995ApJ...449..930Z}, normalised the PDS according to \citet{1983ApJ...272..256L} and converted to square fractional rms \citep{1990A&A...227L..33B}. The PDS were fitted with models composed of zero-centered Lorentzians for band-limited noise (BLN) components, and Lorentzians for quasi-periodic oscillations (QPOs). 

\subsection{\xmm}
An \xmm\ ToO observation of \gs\ was taken on August 6th 2015 (Obs.\,id.: 0727961501) with the EPIC/pn camera in timing mode. The exposure was 11.1 ks. We filtered and extracted the pn event file, using standard SAS (version 14.0.0) tools, paying particular attention to extract the list of photons not randomized in time. After using the SAS task \texttt{epatplot} to investigate whether the observation is affected by pile-up, we selected source photons from two stripes (30$\le$RAWX$\le$35 and 39$\le$RAWX$\le$44) centred on the column with the highest count rate to minimize the effect of pile-up. This selection results in an observed pattern distribution that follows the theoretical prediction quite nicely. We selected single and double events (PATTERN$\le$4) for our study.
 
We produced PDS in different energy bands for the whole observation. As for the \swift\ data, the contribution due to Poissonian noise was subtracted and the normalised PDS were converted to square fractional rms.

\begin{figure}
\resizebox{\hsize}{!}{\includegraphics[clip,angle=0]{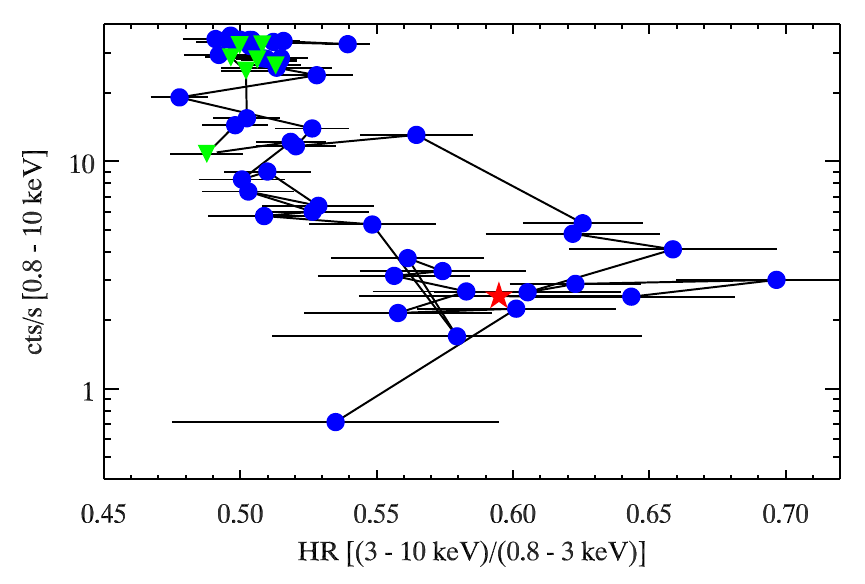}}
\caption{Hardness-intensity diagram, derived using \swift/XRT count rates. Each data point represents one observation. Observations in which a type-C QPO has been detected are marked by (green) triangles, and the (red) star indicates the first observation.}
\label{Fig:HID}
\end{figure}

\begin{figure}
\resizebox{\hsize}{!}{\includegraphics[clip,angle=0]{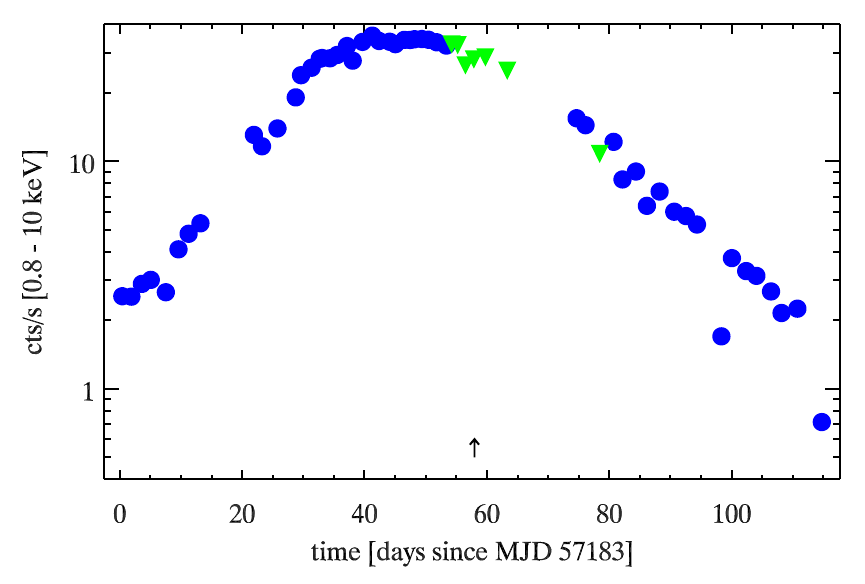}}
\caption{Light curve of the 2015 outburst, based on \swift/XRT count rates. Each data point represents one observation. Observations in which a type-C QPO has been detected are marked by (green) triangles. The arrow marks the date of the \xmm\ ToO observation. T=0 corresponds to June 10th 2015 00:00:00.000 UTC.}
\label{Fig:LC}
\end{figure}

\begin{figure}
\resizebox{\hsize}{!}{\includegraphics[clip,angle=0]{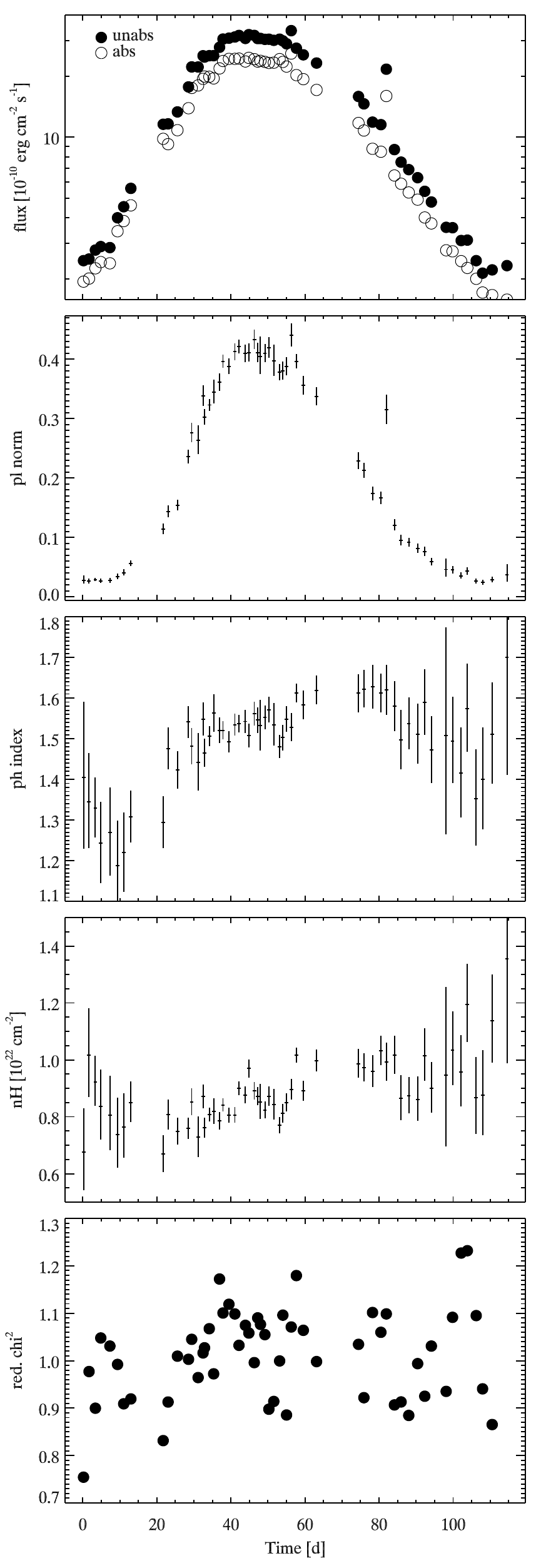}}
\caption{Evolution of spectral parameters.}
\label{Fig:spec}
\end{figure}

\section[]{Results}
\label{Sec:res}
\subsection{\swift\ data}
\label{SubSec:swres}
\subsubsection{Spectral properties}
Using the \swift/XRT data we determined the source count rates in the total (0.8 -- 10 keV), soft (0.8 -- 3 keV), and hard (3 -- 10 keV) energy band, and derived a hardness ratio by dividing the count rate observed in the hard band by the one obtained in the soft band. The HID of the 2015 outburst of \gs\ is shown in Fig.~\ref{Fig:HID} and the light curve in Fig.~\ref{Fig:LC}. After the detection of the outburst the source shows an increase in count rate and the HR softens. This is the ``classical'' behaviour of a BHB observed at the beginning of an outburst. After about  23 days the HR falls below 0.54. In the then following 18 days the count rate increases until the maximum count rate is reached on day 41; then follows a decrease in count rate. During all this time the HR stays around 0.5 (0.47 -- 0.54). After day 94 the HR starts to harden. This shows that \gs\ remains in the hard state during its entire outburst never making it into the soft state. This type of outburst, which is a.\ k.\ a.\ ``failed'' outburst, has also been observed in the previous (1997) outburst of \gs, and in other sources like \h1743\ \citep{2009MNRAS.398.1194C,2016arXiv160201550S}, XTE\,1550--564 \citep{2005ApJ...625..923S}, Aql\,X--1 \citep{2006A&A...451.1045R}, Swift\,J174510.8--262411 \citep{2015arXiv151202805D}, and V404\,Cyg, A\,1542--62, 4U\,1543--475, GRO\,J0422+32, GRO\,J1719--24, GRS1737--21 \citep{2004NewA....9..249B}.    

We used \textsc{Xspec} \citep[V.\ 12.8.2;][]{1996ASPC..101...17A} to fit the energy spectra in the 0.8 -- 10 keV range. Softer energies (below 0.8 keV) are omitted as the spectra are affected by a turn-up in this energy range, which is due to RMF redistribution modelling issues\footnote{http://www.swift.ac.uk/analysis/xrt/digest\_cal.php}. 
Spectra were grouped to contain at least 20 counts in each bin. Grouping data allows us to use $\chi^2$ minimisation to obtain the best fit. The observed spectra can be well described by an absorbed \citep[\texttt{tbabs};][]{2000ApJ...542..914W} power law model. Individual spectral parameters, using the abundances of \citet{2000ApJ...542..914W} and the cross sections given in \citet{1996ApJ...465..487V}, and reduced $\chi^2$ values are given in Table \ref{Tab:Spec_par_sw}. The temporal evolution of the foreground absorption, photon index, and power law normalisation are shown in Fig.~\ref{Fig:spec}. While the photon index shows an increase followed by a decrease during the outburst, it remains hard during the whole outburst. At the beginning of the outburst photon indices in the range of 1.3 -- 1.4 are observed, while at the end photon indices are in the range of 1.4 -- 1.5. Photon indices below 1.6 are unusually low for a BHB. The highest photon indices of about 1.6 are observed during outburst decay about 30 -- 35 days after the maximum luminosity has been reached. The obtained photon indices depend on the observed foreground absorption as both parameters are correlated (\eg\ a higher foreground absorption will lead to a smaller photon index). Using the averaged foreground absorption of $n_H = 8.6\times10^{21}$ cm$^{-2}$, the highest photon indices are reached during outburst rise, about 6 -- 12 days before the maximum luminosity has been reached, and during the remaining outburst the photon indices decrease slowly. Although the evolution of the photon index depends on the assumed foreground absorption (variable versus constant), the range of observed photon indices is the same in both cases.

From spectral fits of RXTE data of the 1997 outburst of \gs\ a photon index of about 1.4 to 1.5 has been obtained \citep{2000ApJ...530..955R}. The hard photon index indicates that the source remained in the hard state during the entire outburst. Using the obtained spectral parameters we converted the highest observed \swift/XRT count rate into an RXTE/PCA count rate using PIMMS. Comparing the count rate to the ones presented in \citet{2001MNRAS.323..517B} for the 1997 outburst, we found that a similar maximum brightness was reached in both outbursts. The evolution of the (un-)absorbed flux during the 2015 outburst obtained from the spectral fits is shown in Fig.~\ref{Fig:spec}. 

\subsubsection{Timing properties}
In general the PDS show a BLN component. For observations taken during outburst rise a peaked noise component with a characteristic frequency \citep[$\nu_{\mr{max}}=\sqrt{\nu^2+\Delta^2}$, where $\nu$ is the centroid frequency, and $\Delta$ is the half width at half maximum][]{2002ApJ...572..392B} in the range of 0.03 -- 0.15 Hz is present. The overall evolution of the characteristic frequency of this feature is an increase with ongoing outburst. During the entire outburst the observed fractional rms was higher than 10 per cent, which confirms that the source stayed in the hard state, as the transition to the soft intermediate state happens at fractional rms below 10 per cent \citep{2011MNRAS.410..679M}. For observations taken on day 54, 55, 56, 58, 60, 63, and 78 of the outburst a type-C QPO \citep{1999ApJ...526L..33W,2005ApJ...629..403C} is present (parameters can be found in Table~\ref{Tab:sw_qpo}). 
All these observations are taken after the outburst reached its maximum (Fig.~\ref{Fig:LC}). The characteristic frequency of the QPO is in the range of 0.17 -- 0.25 Hz, and the highest frequency is observed in the observation that shows the first peak in the light curve after the outburst maximum. The obtained Q values ($Q=\nu/(2\Delta)$) of these QPOs are $\ge4.5$. For the observations taken on day 55 and 58 an upper harmonic (day 55: Q$_{\mr{uh}}$=5.4, $\nu_{\mr{max;uh}}=0.42\pm0.02$ Hz; $\nu_{\mr{max;qpo}}=0.18\pm0.01$ Hz; day 58: Q$_{\mr{uh}}>$47.5, $\nu_{\mr{max;uh}}=0.399^{+0.002}_{-0.001}$ Hz; $\nu_{\mr{max;qpo}}=0.193^{+0.006}_{-0.007}$ Hz) is present, while the observation taken on day 63 shows a lower harmonic (Q$_{\mr{lh}}$=5.2, $\nu_{\mr{max;lh}}=0.087^{+0.005}_{-0.006}$ Hz; $\nu_{\mr{max;qpo}}=0.195\pm0.007$ Hz).

\begin{table}
\caption{Parameters of the PDS in the \xmm\ observation}
\begin{center}
\begin{tabular}{lrr}
\hline\noalign{\smallskip}
 \multicolumn{1}{c}{Parameter} & \multicolumn{1}{c}{1 -- 2 keV} & \multicolumn{1}{c}{2 -- 10 keV} \\
\hline\noalign{\smallskip}
rms$_{\mr{BLN}}$ [\%] & $14.3^{+0.8}_{-0.9}$ & $16.5\pm0.7$ \\
\smallskip
$\nu_{\mr{BLN}}$ [Hz] & $2.85^{+0.85}_{-0.66}$ & $3.20^{+0.58}_{-0.52}$ \\
\smallskip
$\nu_{\mr{QPO}}$ [Hz]  & $0.191\pm0.002$ & $0.190^{+0.001}_{-0.002}$ \\
\smallskip
$\Delta_{\mr{QPO}}$ [Hz] & $0.022\pm0.004$ & $0.016\pm0.003$ \\
\smallskip
rms$_{\mr{QPO}}$ [\%] & $11.8^{+0.8}_{-0.9}$ & $13.9\pm0.8$ \\
\smallskip
$\nu_{\mr{PN1}}$ [Hz] & $0.43^{+0.01}_{-0.02}$ & $0.40\pm0.02$ \\
\smallskip
$\Delta_{\mr{PN1}}$ [Hz] & $0.18^{+0.03}_{-0.02}$ & $0.26\pm0.03$ \\
\smallskip
rms$_{\mr{PN1}}$  [\%] & $13.7^{+0.9}_{-1.0}$ & $18.2^{+1.0}_{-1.1}$\\
\smallskip
$\nu_{\mr{PN2}}$ [Hz] & $0.074^{+0.007}_{-0.006}$ & $0.082^{+0.006}_{-0.005}$ \\
\smallskip
$\Delta_{\mr{PN2}}$ [Hz] & $0.084^{+0.017}_{-0.013}$ & $0.077^{+0.013}_{-0.010}$ \\
\smallskip
rms$_{\mr{PN2}}$ [\%] & $15.5^{+1.2}_{-0.9}$ & $17.5^{+1.2}_{-0.9}$ \\
\hline\noalign{\smallskip} 
\end{tabular} 
\end{center}
\label{Tab:bbPDS}
\end{table}

\begin{figure}
\resizebox{\hsize}{!}{\includegraphics[clip,angle=0]{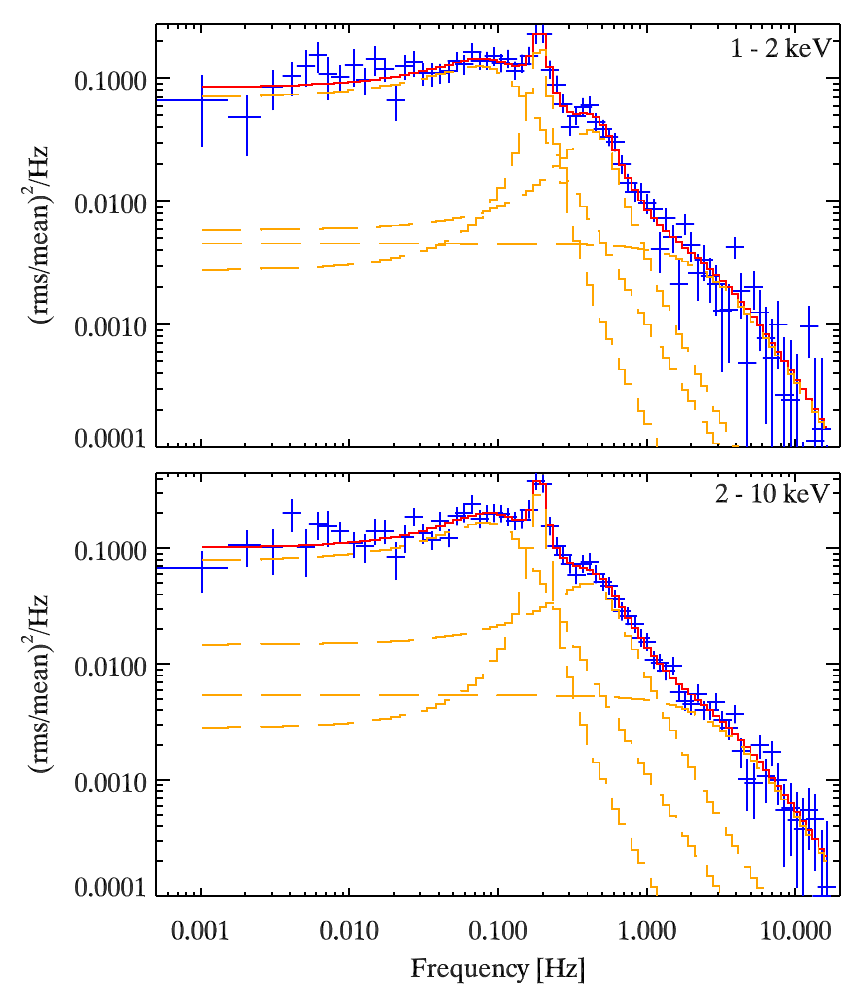}}
\caption{Power density spectra in the 1 -- 2 (upper panel) and 2 -- 10 keV range (lower panel) fitted with Lorentzians.}
\label{Fig:pds}
\end{figure}

\begin{figure}
\resizebox{\hsize}{!}{\includegraphics[clip,angle=0]{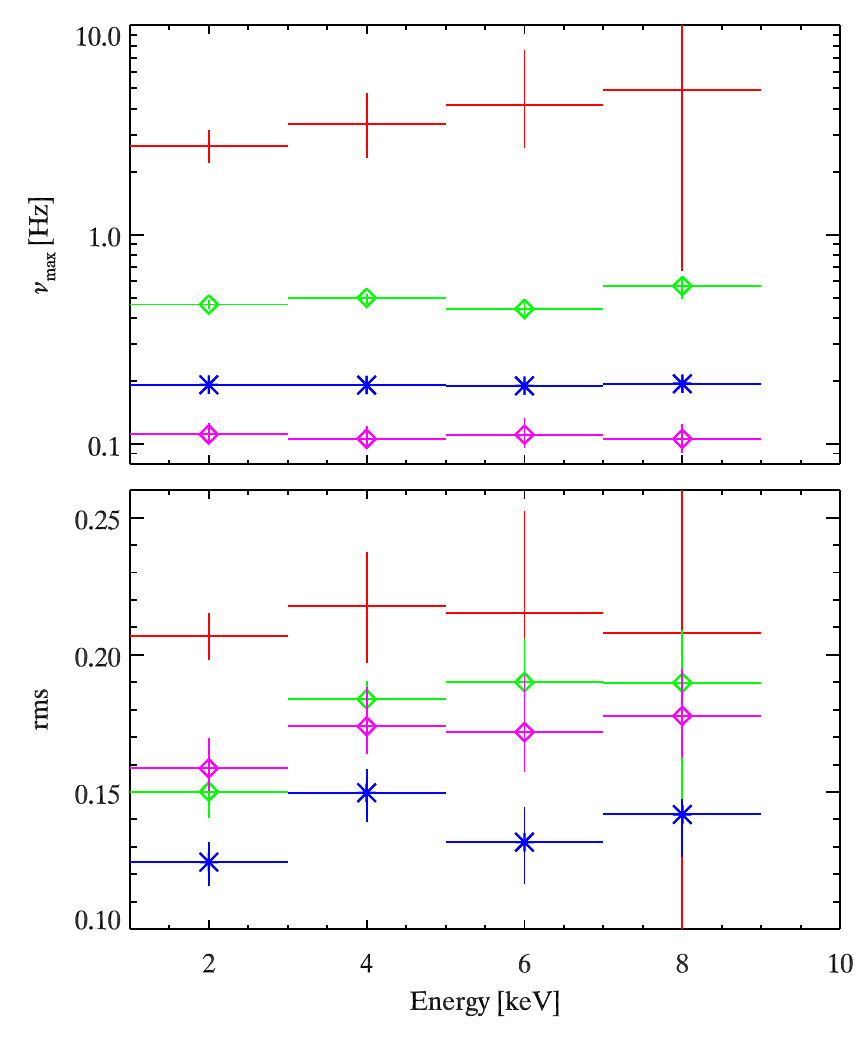}}
\caption{Energy dependence of the characteristic frequency (upper panel) and rms spectra of the BLN (red cross), QPO (blue X), upper peaked noise (green circle), and lower peaked noise (magenta diamond).}
\label{Fig:rms_spec_indv}
\end{figure}

\begin{figure}
\resizebox{\hsize}{!}{\includegraphics[clip,angle=0]{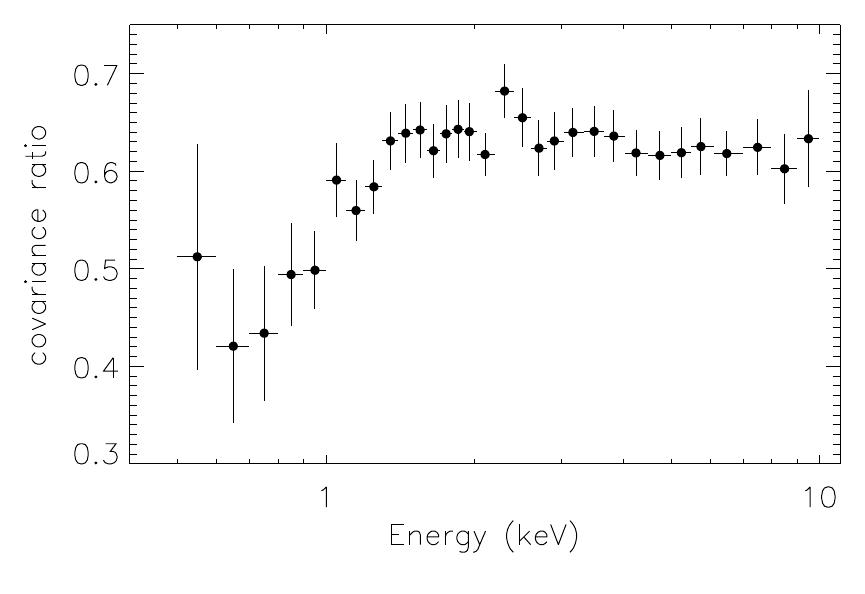}}
\caption{Covariance ratio, derived by dividing the long timescale covariance spectrum by the short timescale one.}
\label{Fig:covspecratio}
\end{figure}

\begin{figure}
\resizebox{\hsize}{!}{\includegraphics[clip,angle=0]{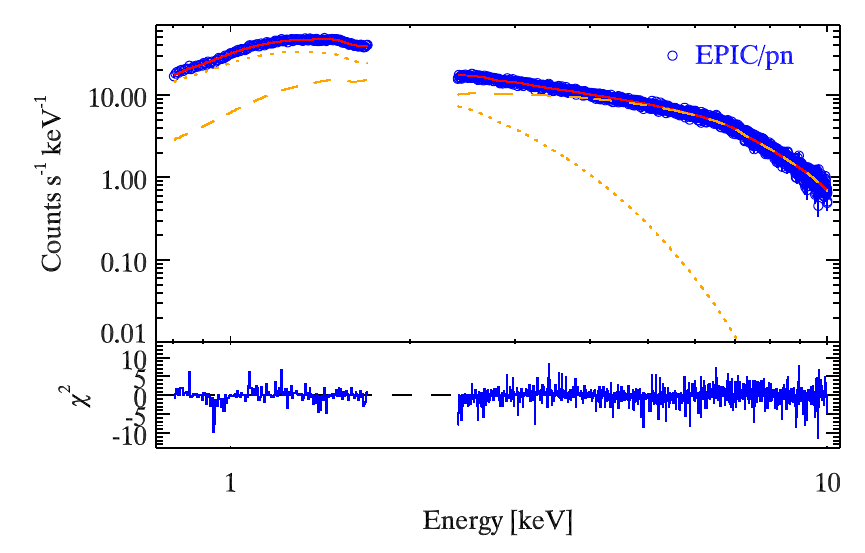}}
\caption{Energy spectrum of the \xmm\ observation based on EPIC/pn data in the range of 0.8 -- 10 keV, ignoring energies between 1.7 and 2.4 keV. Data points, best fit model (model B; see Table~\ref{Tab:Spec_par}) and individual model components are given (disc blackbody: dotted line; Comptonization component: dashed line).}
\label{Fig:espec}
\end{figure}

\begin{table*}
\caption{Spectral parameters obtained by fitting the EPIC/pn spectrum.}
\begin{center}
\begin{tabular}{lllll}
\hline\noalign{\smallskip}
 \multicolumn{1}{c}{component} & \multicolumn{1}{c}{model A}  & \multicolumn{1}{c}{model B}  & \multicolumn{1}{c}{model C}& \multicolumn{1}{c}{model D}\\
 \hline\noalign{\smallskip}
N$_{\mr{H}}$ [$10^{21}$ cm$^{-2}$]&$8.36^{+0.05}_{-0.04}$ & $6.76\pm0.07$ & $7.04^{+0.07}_{-0.05}$ &$8.77^{+0.20}_{-0.19}$\\
\noalign{\smallskip}
R$_{\mr{in}} [\mr{km/}\sqrt{\cos(\theta)}]^{\ddagger}$& &$55.6^{+5.2}_{-3.6}$ &$55.1^{+5.6}_{-3.7}$ &$105.3^{+19.4}_{-30.5}$\\
\noalign{\smallskip}
T$_{\mr{in}}$ [keV]& &$0.62_{-0.04}^{+0.03}$ & $0.50^{+0.01}_{-0.02}$ & $0.33\pm0.02$\\
\noalign{\smallskip}
$\Gamma$ &$1.577^{+0.004}_{-0.003}$ &$1.44^{+0.06}_{-0.05}$ & $1.51^{+0.05}_{-0.03}$ & $1.52\pm0.01$\\
\noalign{\smallskip}
kT$_{\mr{e}}$ [keV]& &$2.4\pm0.1$ &$10.2_{-3.7}^{+5.0}$ &\\
\noalign{\smallskip}
N$_{\mr{compton}}$ & $0.602\pm0.003$& $0.10^{+0.03}_{-0.01}$ & $0.40\pm0.01$ & $0.41_{-0.03}^{+0.01}$\\
\noalign{\smallskip}
rel.~reflection & &$0.64^{+0.18}_{-0.15}$ & & $0.39^{+0.07}_{-0.04}$\\
\noalign{\smallskip}
E$_{\mr{cutoff}}$&& &$6.82\pm0.08$  \\
\noalign{\smallskip}
E$_{\mr{fold}}$&& & $9.4\pm0.5$\\
\noalign{\smallskip}
a && & &$-0.998 - -0.930$\\
\noalign{\smallskip}
log$\xi$&& & & $3.30_{-0.04}^{+0.02}$\\
\noalign{\smallskip}
Z$_{\mr{Fe}}$ [Z$_{\mr{Fe},\odot}$]&& & & $5.0\pm0.5$\\
\noalign{\smallskip}
$\chi^2/$dof & 2219.0/1701& 1682.1/1697  &1638.4/1696 & 1595.7/1695\\
\hline\noalign{\smallskip} 
\end{tabular} 
\end{center}
\normalsize
Notes: \\
$^{\ddagger}$: assuming a distance of 25 kpc  \citep{2009ApJS..181..238C}\\
model A: \texttt{TBabs $\times$ powerlaw}\\
model B: \texttt{TBabs $\times$ (diskbb + reflect(nthcomp))}\\
model C: \texttt{TBabs $\times$ (diskbb + highecut $\times$ nthcomp)}\\
model D: \texttt{TBabs $\times$ (diskbb + relxill)}
\label{Tab:Spec_par}
\end{table*}

\subsection{\xmm\ data}
\label{SubSec:xmmres}
\subsubsection{Timing properties}
The soft band PDS (1 -- 2 keV) can be well fitted by one zero centered Lorentzian describing the BLN component, one peaked Lorentzian for the QPO at $0.192\pm0.002$ Hz and two peaked Lorentzians for two noise components that peak at $0.11^{+0.02}_{-0.01}$ and $0.47^{+0.02}_{-0.03}$ Hz (Fig.~\ref{Fig:pds}). The broadband PDS (2 -- 10 keV) can be fitted with the same components (see Table~\ref{Tab:bbPDS}). To study the energy dependence of the rms amplitude and of the characteristic frequency we fit PDS obtained in energy bands with a width of 2~keV, namely: 1 -- 3, 3 -- 5, 5 --7, and 7 -- 9 keV. The characteristic frequency of the BLN component (red crosses in upper panel of Fig.~\ref{Fig:rms_spec_indv}) shows a slight increase with increasing energy. This energy dependence of the characteristic frequency has already been observed during LHS observations of other BHBs \citep{2015MNRAS.452.3666S}. 

We also derive covariance ratios by dividing the long timescale covariance spectrum by the short timescale one. The covariance spectra are derived following \citet{2009MNRAS.397..666W} and \citet{2015MNRAS.452.3666S}. Taking a look at the PDS (Fig.~\ref{Fig:pds}) we use bins of 8 times the frame time measured in segments of 30 bins for the short timescale range and bins of 646 times the frame time measured in segments of 70 bins for the long timescale range. With this selection we compare variability measured in the flat part of the PDS to the one measured in the decaying part, excluding contribution from the QPO. The obtained covariance ratios are shown in Fig.~\ref{Fig:covspecratio}. At energies above 1 keV the covariance ratios are flat, while at lower energies the covariance ratio decreases with decreasing energy. This behaviour clearly differs from the increase of the covariance ratio towards lower energies, which has been found in later phases of the LHS in \gx339\ and Swift\,J1753.5-0125 and has been interpreted as the sign of additional disc variability on longer timescales \citep{2009MNRAS.397..666W}. It is more consistent with the flat covariance ratio observed in the 2008 and 2014 outbursts of \h1743\ \citep{2016arXiv160201550S}.

\subsubsection{Spectral properties}

We fit the averaged EPIC/pn spectrum within \textsc{isis} V.~1.6.2 \citep{2000ASPC..216..591H} in the 0.8 -- 10 keV range, grouping the data to ensure that we have at least 20  source counts in each bin. We exclude the energy range between 1.7 and 2.4 keV from the EPIC/pn spectrum, as this energy range is affected by features caused by gain shift due to Charge-transfer inefficiency around 1.8 and 2.2 keV. We include a systematic uncertainty of 1 per cent. Using an absorbed power law model, as we did for the \swift\ data, and including an additional Gaussian component to model the excess emission below 1.3 keV \citep{2011MNRAS.411..137H,2014A&A...571A..76D}, we obtain a foreground absorption of $8.36^{+0.05}_{-0.04}\times10^{21}$ cm$^{-2}$ and a photon index of $1.577_{-0.003}^{+0.004}$, which are lower than the values obtained from the \swift\ observation taken on the same day (n$_{\mr{H}}=10.17\pm0.26\times10^{21}$ cm$^{-2}$; $\Gamma=1.612\pm0.023$). With a reduced $\chi^2$ of 2219.0/1701 the fit is unacceptable, indicating that additional components are needed to describe the EPIC/pn spectrum properly. 

Using an absorbed (\texttt{tbabs}) disc blackbody \citep[\texttt{diskbb};][]{1984PASJ...36..741M} plus a thermal Comptonisation component \citep[\texttt{nthcomp};][]{1996MNRAS.283..193Z,1999MNRAS.309..561Z} including reflection \citep[\texttt{reflect};][]{1995MNRAS.273..837M} we got an acceptable fit with $\chi^2_{\mr{red}}=$ 1682.1/1697 (Fig.~\ref{Fig:espec}). The individual  spectral parameters are given in Table~\ref{Tab:Spec_par}. Including a disc blackbody and reflection, the photon index gets even lower ($\Gamma=1.44^{+0.06}_{-0.05}$). The obtained inner disc temperature of $0.62_{-0.04}^{+0.03}$ keV is lower than what has been observed with RXTE for a hard state \citep[$kT_{\mr{in}}\sim 0.8 - 0.9$ keV; see \eg\ ][]{2010MNRAS.408.1796M,2011MNRAS.418.2292M,2011MNRAS.415..292M}. The inner disc radius is at least $55.6^{+5.2}_{-3.6}$ km/$\sqrt{\cos(\theta)}$ (assuming a distance of more than 25 kpc), which is bigger than what has been observed with RXTE in other BHBs \citep[$R_{\mr{in}}\sim 40$ km; see \eg\ ][]{2010MNRAS.408.1796M,2011MNRAS.418.2292M,2011MNRAS.415..292M}. This can be related to the fact that RXTE allowed only to get energy spectra above 3 keV, in the energy range that is dominated by non-thermal emission in the LHS, and therefore no disc component was need to fit these spectra.

To compare the energy spectrum to those obtained in \citet{2015MNRAS.452.3666S} we also fit the model used in this paper, which includes a high-energy cutoff component (\texttt{highecut}) instead of a reflection component. We got an acceptable fit with $\chi^2_{\mr{red}}=$1638.4/1696. The photon index ($\Gamma=1.51^{+0.05}_{-0.03}$) and inner disc radius (R$_{\mr{in}}=55.1^{+5.6}_{-3.7}$ km/$\sqrt{\cos(\theta)}$) are in agreement with the values obtained with the model including a reflection component, while the inner disc temperature (T$_{\mr{in}}=0.50^{+0.01}_{-0.02}$ keV) is slightly lower.  

In addition, we substitute the reflected non-thermal component by the relativistic reflection model \texttt{relxill} \citep{2014MNRAS.444L.100D,2014ApJ...782...76G} and obtain a fit with $\chi^2_{\mr{red}}$ of 1595.7/1695. With this model we obtain a photon index of $1.52\pm0.01$, an inner disc temperature of $0.33\pm0.02$ keV, and an inner disc radius of $105.3^{+19.4}_{-30.5}$ km/$\sqrt{\cos(\theta)}$. The \texttt{relxill} model gives a spin between $-0.930$ and $-0.998$. 

\section[]{Discussion}
\label{Sec:dis}
We used \swift/XRT and \xmm/EPIC pn data to study the 2015 outburst of \gs. The spectra obtained from the \swift\ monitoring data can be fitted with a power law model with a photon index in the range of 1.3 -- 1.6. This indicates that the source stays in the hard state during the entire outburst. This result is confirmed by the timing properties. 

Before the 2015 outburst \gs\ showed an outburst in 1997 during which it also stayed in the LHS during the entire outburst \citep{2000ApJ...530..955R,2001MNRAS.323..517B}. During the 1997 outburst a similar photon index was observed \citep[$\Gamma=1.4-1.5$;][]{2000ApJ...530..955R}. There are three even earlier detections of X-ray activity of transient sources that can be spatially related to \gs. During two of these detections the source was in the HSS, while the remaining one caught a source in the LHS  \citep[see][and references therein]{1995xrbi.nasa..126T}. In case all observations are related to the same source this source shows ``normal'' outbursts with LHS and HSS and so-called ``failed'' outbursts where it stays in the LHS during the entire outburst. A BHB that is known to show both types of outbursts is \h1743\ \citep[][and references therein]{2016arXiv160201550S}. Unlike \gs\ \h1743\ shows outbursts quite frequently. Another possibility is that only the ``failed'' outbursts are related with \gs\ and that the HSS are from a nearby transient source. About ten X-ray binaries (including neutron stars and black holes) are known that have only been observed during outbursts where they stayed in the LHS \citep[][and references therein]{2009MNRAS.398.1194C}. To be certain that \gs\ belongs to the type of BHBs that show ``normal'' outbursts we would need to observe a HSS with a modern X-ray satellite that allows to confirm that the thermal X-ray emission is spatially consistent with the position of \gs\ obtained during the observation of the Comptonized emission during the 2015 outburst.

The covariance ratio obtained from the \xmm\ observation decreases with decreasing energy below 1 keV, while the one obtained for LHS observations of GX\,339--4 and Swift\,J1753.5--0125 increases with decreasing energy \citep{2009MNRAS.397..666W,2015MNRAS.452.3666S}. The increase of the covariance ratio towards lower energies has been interpreted as a sign of additional disc variability on longer timescales. Thus the decrease observed in \gs\ can be either regarded as a sign of missing disc variability on longer timescales or as a sign of additional variability on short timescales. For two observations of \h1743\ taken during the 2008 and 2014 ``failed'' outbursts we also found covariance ratios that do not show an increase at lower energies. In case of \h1743\ the covariance ratios remained rather flat \citep{2016arXiv160201550S}.

In \citet{2015MNRAS.452.3666S} we investigated the energy dependence of the characteristic frequency of the band-limited noise in PDS and found that for a noise component with a characteristic frequency above 1 Hz the characteristic frequency in the 1 -- 2 keV band is lower than in the 4 -- 8 keV. From the \xmm\ data of \gs\ we obtain a characteristic frequency of $2.85^{+0.85}_{-0.66}$ Hz in the 1 -- 2 keV band which is lower than the characteristic frequency of $4.64^{+1.53}_{-1.14}$Hz in the 4 -- 8 keV band, although this result is not significant as the errors are big. The sources investigated in \citet{2015MNRAS.452.3666S} showed either an increase of the covariance ratio towards lower energies or the covariance ratio was flat. The fact that we observe the same energy dependence of the characteristic frequency as in \citet{2015MNRAS.452.3666S} while the energy dependence of the covariance ratio at soft energies is inverted gives additional evidence that there is no connection between the energy dependence of the characteristic frequency and the covariance ratio. In the cases of \h1743\ where we found flat covariance ratios there is no energy dependence of the characteristic frequency \citep{2015MNRAS.452.3666S,2016arXiv160201550S}.

Fitting the energy spectrum with the same model that has been used in \citet{2015MNRAS.452.3666S}, we found that the energy spectrum of \gs\ differed significantly from those studied in \citet{2015MNRAS.452.3666S}. The inner disc temperature of \gs\ is significantly higher than the temperatures found in the previous study, while the disc blackbody normalisation, photon index, and cut-off and fold energies are lower. With a higher inner disc temperature and a smaller inner disc radius the observed covariance ratio cannot be explained by a faint disc component and it is more likely that the differences in covariance ratio are related to some changes in the accretion process. We note that all three observations that do not show increasing covariance ratio towards lower energies have energy spectra that require a rather low photon index and were taken during ``failed'' outbursts. Therefore different shapes of covariance ratio, although observed at soft energies, might be driven by changes in the Comptonizing component or they indicate changes in the accretion geometry that determine if a BHB goes into a ``normal'' or ``failed'' outburst. In the case of \h1743\ the different shape of the covariance ratio can also be related to the higher inclination angle of this source in comparison to the inclination of GX\,339--4 or Swift\,J1753.5-0125 \citep{2016arXiv160201550S}. For \gs\ the inclination angle is not known. 

A further investigation of these different possibilities must be the aim of future studies as more data are needed. Further insight can be obtained by observations of other sources during a ``failed'' outburst or at high inclination to extent the size of the sample or by an observation of \h1743\ in an early LHS with \xmm\ during a ``normal'' outburst.

\section*{Acknowledgments}
This project is supported by the Ministry of Science and Technology of
the Republic of China (Taiwan) through grants 103-2628-M-007-003-MY3
and 104-281-M-007-060.

\bibliographystyle{mn2e}
\bibliography{biblio.bib} 

\begin{table}
\caption{\label{Tab:Spec_par_sw}Spectral parameters obtained by fitting the \swift/XRT spectra.}
\begin{center}
\begin{tabular}{rrrlllrrrllrr}
\hline\noalign{\smallskip}
obs$^{\dagger}$&   time$^{*}$&  $\chi^2_{\mr{red}}$ & n$_H^{\#}$                      &  $\Gamma$                    &  norm$^{+}$                     & flux$_{\mr{abs}}^{\ddagger}$     & flux$_{\mr{unabs}}^{\ddagger}$ &  $\chi^{2 \spadesuit}_{\mr{red}}$ &  $\Gamma^{\spadesuit}$                    &  norm$^{+,\spadesuit}$                     & flux$_{\mr{abs}}^{\ddagger,\spadesuit}$     & flux$_{\mr{unabs}}^{\ddagger,\spadesuit}$ \\
 \hline\noalign{\smallskip}
02 &   0.4 & 0.75 & $ 6.8_{-1.3}^{+1.5}$  & $1.40_{-0.18}^{+0.19}$ & $0.028_{-0.006}^{+0.008}$ &    1.93 &  2.45 & 0.76 & $1.31_{-0.10}^{+0.10}$ & $0.024_{-0.002}^{+0.002}$ &  1.98 &  2.41 \\
\smallskip
03 &   1.9 & 0.98 & $10.2_{-1.5}^{+1.6}$  & $1.35_{-0.11}^{+0.12}$ & $0.026_{-0.004}^{+0.005}$ &    2.00 &  2.50 & 1.00 & $1.25_{-0.06}^{+0.06}$ & $0.023_{-0.002}^{+0.002}$ &  2.05 &  2.45 \\
\smallskip
04 &   5.1 & 1.05 & $ 8.4_{-1.2}^{+1.3}$  & $1.24_{-0.10}^{+0.10}$ & $0.027_{-0.003}^{+0.004}$ &    2.42 &  2.88 & 1.04 & $1.26_{-0.06}^{+0.06}$ & $0.027_{-0.002}^{+0.002}$ &  2.41 &  2.89 \\
\smallskip
05 &   3.6 & 0.90 & $ 9.2_{-0.8}^{+0.9}$  & $1.33_{-0.07}^{+0.07}$ & $0.029_{-0.003}^{+0.003}$ &    2.25 &  2.77 & 0.90 & $1.29_{-0.04}^{+0.04}$ & $0.027_{-0.001}^{+0.001}$ &  2.27 &  2.75 \\
\smallskip
06 &   7.5 & 1.03 & $ 8.1_{-1.2}^{+1.4}$  & $1.27_{-0.11}^{+0.11}$ & $0.027_{-0.004}^{+0.004}$ &    2.38 &  2.85 & 1.03 & $1.31_{-0.06}^{+0.06}$ & $0.029_{-0.002}^{+0.002}$ &  2.36 &  2.86 \\
\smallskip
07 &   9.6 & 0.99 & $ 7.4_{-1.2}^{+1.3}$  & $1.19_{-0.11}^{+0.11}$ & $0.034_{-0.005}^{+0.005}$ &    3.43 &  4.00 & 1.01 & $1.27_{-0.07}^{+0.07}$ & $0.039_{-0.003}^{+0.003}$ &  3.35 &  4.04 \\
\smallskip
08 &  11.3 & 0.91 & $ 7.6_{-1.1}^{+1.2}$  & $1.22_{-0.10}^{+0.10}$ & $0.040_{-0.005}^{+0.006}$ &    3.85 &  4.53 & 0.92 & $1.29_{-0.06}^{+0.06}$ & $0.045_{-0.003}^{+0.003}$ &  3.78 &  4.57 \\
\smallskip
09 &  13.2 & 0.92 & $ 8.5_{-0.7}^{+0.7}$  & $1.31_{-0.06}^{+0.07}$ & $0.056_{-0.005}^{+0.005}$ &    4.60 &  5.58 & 0.92 & $1.32_{-0.04}^{+0.04}$ & $0.057_{-0.002}^{+0.002}$ &  4.59 &  5.59 \\
\smallskip
10 &  21.9 & 0.83 & $ 6.7_{-0.6}^{+0.7}$  & $1.29_{-0.06}^{+0.06}$ & $0.114_{-0.009}^{+0.010}$ &    9.80 & 11.56 & 0.91 & $1.44_{-0.04}^{+0.04}$ & $0.139_{-0.006}^{+0.006}$ &  9.43 & 11.84 \\
\smallskip
11 &  23.2 & 0.91 & $ 8.1_{-0.5}^{+0.5}$  & $1.48_{-0.05}^{+0.05}$ & $0.144_{-0.009}^{+0.010}$ &    9.22 & 11.62 & 0.92 & $1.52_{-0.03}^{+0.03}$ & $0.153_{-0.005}^{+0.005}$ &  9.12 & 11.74 \\
\smallskip
12 &  25.8 & 1.01 & $ 7.5_{-0.5}^{+0.5}$  & $1.42_{-0.05}^{+0.05}$ & $0.154_{-0.009}^{+0.010}$ &   10.81 & 13.31 & 1.04 & $1.51_{-0.03}^{+0.03}$ & $0.175_{-0.005}^{+0.005}$ & 10.56 & 13.56 \\
\smallskip
14 &  28.8 & 1.00 & $ 7.6_{-0.4}^{+0.4}$  & $1.54_{-0.04}^{+0.04}$ & $0.236_{-0.011}^{+0.012}$ &   13.88 & 17.68 & 1.04 & $1.62_{-0.02}^{+0.02}$ & $0.264_{-0.006}^{+0.006}$ & 13.58 & 18.08 \\
\smallskip
15 &  29.6 & 1.05 & $ 8.5_{-0.5}^{+0.5}$  & $1.48_{-0.04}^{+0.05}$ & $0.276_{-0.015}^{+0.017}$ &   17.47 & 22.19 & 1.04 & $1.49_{-0.03}^{+0.03}$ & $0.279_{-0.007}^{+0.008}$ & 17.43 & 22.23 \\
\smallskip
17 &  31.4 & 0.96 & $ 7.3_{-0.7}^{+0.7}$  & $1.44_{-0.07}^{+0.07}$ & $0.263_{-0.023}^{+0.025}$ &   17.98 & 22.17 & 1.00 & $1.55_{-0.04}^{+0.04}$ & $0.304_{-0.013}^{+0.013}$ & 17.50 & 22.71 \\
\smallskip
18 &  32.7 & 1.02 & $ 8.7_{-0.4}^{+0.4}$  & $1.55_{-0.04}^{+0.04}$ & $0.338_{-0.017}^{+0.018}$ &   19.37 & 25.18 & 1.01 & $1.54_{-0.02}^{+0.02}$ & $0.335_{-0.008}^{+0.008}$ & 19.40 & 25.13 \\
\smallskip
19 &  33.1 & 1.03 & $ 7.6_{-0.3}^{+0.3}$  & $1.46_{-0.03}^{+0.03}$ & $0.303_{-0.013}^{+0.014}$ &   19.89 & 24.81 & 1.07 & $1.54_{-0.02}^{+0.02}$ & $0.338_{-0.007}^{+0.007}$ & 19.49 & 25.27 \\
\smallskip
20 &  34.4 & 1.07 & $ 8.1_{-0.2}^{+0.2}$  & $1.51_{-0.02}^{+0.03}$ & $0.323_{-0.010}^{+0.010}$ &   19.83 & 25.21 & 1.09 & $1.55_{-0.02}^{+0.02}$ & $0.343_{-0.005}^{+0.005}$ & 19.61 & 25.48 \\
\smallskip
21 &  35.5 & 0.97 & $ 8.2_{-0.4}^{+0.5}$  & $1.56_{-0.05}^{+0.05}$ & $0.344_{-0.019}^{+0.021}$ &   19.47 & 25.22 & 0.98 & $1.60_{-0.03}^{+0.03}$ & $0.362_{-0.010}^{+0.010}$ & 19.29 & 25.46 \\
\smallskip
23 &  37.1 & 1.17 & $ 7.8_{-0.3}^{+0.3}$  & $1.52_{-0.03}^{+0.03}$ & $0.361_{-0.014}^{+0.015}$ &   21.82 & 27.75 & 1.20 & $1.58_{-0.02}^{+0.02}$ & $0.394_{-0.007}^{+0.008}$ & 21.47 & 28.19 \\
\smallskip
24 &  38.1 & 1.10 & $ 8.4_{-0.2}^{+0.2}$  & $1.52_{-0.02}^{+0.02}$ & $0.396_{-0.011}^{+0.011}$ &   23.73 & 30.45 & 1.10 & $1.54_{-0.01}^{+0.01}$ & $0.406_{-0.005}^{+0.005}$ & 23.63 & 30.57 \\
\smallskip
25 &  39.7 & 1.12 & $ 8.0_{-0.3}^{+0.3}$  & $1.49_{-0.03}^{+0.03}$ & $0.388_{-0.013}^{+0.013}$ &   24.31 & 30.76 & 1.14 & $1.54_{-0.02}^{+0.02}$ & $0.413_{-0.007}^{+0.007}$ & 24.03 & 31.10 \\
\smallskip
26 &  41.3 & 1.10 & $ 8.1_{-0.3}^{+0.3}$  & $1.53_{-0.03}^{+0.03}$ & $0.412_{-0.014}^{+0.014}$ &   24.33 & 31.18 & 1.12 & $1.58_{-0.02}^{+0.02}$ & $0.438_{-0.007}^{+0.007}$ & 24.06 & 31.55 \\
\smallskip
27 &  42.4 & 1.03 & $ 9.0_{-0.2}^{+0.2}$  & $1.54_{-0.02}^{+0.02}$ & $0.421_{-0.012}^{+0.012}$ &   24.39 & 31.73 & 1.04 & $1.51_{-0.01}^{+0.01}$ & $0.404_{-0.005}^{+0.005}$ & 24.56 & 31.50 \\
\smallskip
28 &  44.1 & 1.07 & $ 8.8_{-0.3}^{+0.3}$  & $1.54_{-0.03}^{+0.03}$ & $0.409_{-0.015}^{+0.015}$ &   23.61 & 30.66 & 1.07 & $1.53_{-0.02}^{+0.02}$ & $0.403_{-0.007}^{+0.007}$ & 23.67 & 30.57 \\
\smallskip
29 &  45.0 & 1.06 & $ 9.7_{-0.3}^{+0.3}$  & $1.51_{-0.03}^{+0.03}$ & $0.411_{-0.015}^{+0.016}$ &   24.59 & 32.02 & 1.11 & $1.43_{-0.02}^{+0.02}$ & $0.366_{-0.006}^{+0.006}$ & 25.10 & 31.43 \\
\smallskip
30 &  46.5 & 1.00 & $ 8.9_{-0.3}^{+0.3}$  & $1.56_{-0.03}^{+0.03}$ & $0.433_{-0.016}^{+0.017}$ &   24.23 & 31.74 & 1.00 & $1.54_{-0.02}^{+0.02}$ & $0.419_{-0.007}^{+0.007}$ & 24.37 & 31.54 \\
\smallskip
31 &  47.4 & 1.09 & $ 8.7_{-0.3}^{+0.3}$  & $1.55_{-0.03}^{+0.03}$ & $0.410_{-0.016}^{+0.017}$ &   23.57 & 30.63 & 1.09 & $1.54_{-0.02}^{+0.02}$ & $0.406_{-0.007}^{+0.007}$ & 23.62 & 30.57 \\
\smallskip
32 &  48.2 & 1.08 & $ 8.5_{-0.6}^{+0.6}$  & $1.53_{-0.06}^{+0.06}$ & $0.405_{-0.030}^{+0.033}$ &   23.77 & 30.67 & 1.07 & $1.54_{-0.04}^{+0.04}$ & $0.409_{-0.015}^{+0.015}$ & 23.72 & 30.73 \\
\smallskip
33 &  49.4 & 1.05 & $ 8.2_{-0.3}^{+0.3}$  & $1.55_{-0.03}^{+0.03}$ & $0.410_{-0.015}^{+0.016}$ &   23.49 & 30.36 & 1.06 & $1.58_{-0.02}^{+0.02}$ & $0.427_{-0.007}^{+0.007}$ & 23.32 & 30.61 \\
\smallskip
34 &  50.5 & 0.90 & $ 8.7_{-0.3}^{+0.3}$  & $1.57_{-0.03}^{+0.03}$ & $0.419_{-0.017}^{+0.018}$ &   23.21 & 30.40 & 0.90 & $1.56_{-0.02}^{+0.02}$ & $0.415_{-0.008}^{+0.008}$ & 23.25 & 30.34 \\
\smallskip
35 &  51.8 & 0.91 & $ 8.4_{-0.5}^{+0.5}$  & $1.53_{-0.05}^{+0.05}$ & $0.397_{-0.025}^{+0.027}$ &   23.29 & 30.01 & 0.91 & $1.55_{-0.03}^{+0.03}$ & $0.406_{-0.012}^{+0.012}$ & 23.19 & 30.13 \\
\smallskip
36 &  53.4 & 1.00 & $ 7.7_{-0.3}^{+0.3}$  & $1.48_{-0.03}^{+0.03}$ & $0.378_{-0.013}^{+0.014}$ &   24.24 & 30.41 & 1.04 & $1.55_{-0.02}^{+0.02}$ & $0.419_{-0.007}^{+0.007}$ & 23.79 & 30.95 \\
\smallskip
37 &  54.2 & 1.10 & $ 8.1_{-0.3}^{+0.3}$  & $1.50_{-0.03}^{+0.03}$ & $0.380_{-0.015}^{+0.016}$ &   23.48 & 29.83 & 1.11 & $1.54_{-0.02}^{+0.02}$ & $0.402_{-0.008}^{+0.008}$ & 23.25 & 30.12 \\
\smallskip
38 &  55.2 & 0.89 & $ 8.5_{-0.3}^{+0.3}$  & $1.55_{-0.03}^{+0.03}$ & $0.388_{-0.015}^{+0.016}$ &   22.27 & 28.86 & 0.88 & $1.56_{-0.02}^{+0.02}$ & $0.393_{-0.007}^{+0.007}$ & 22.22 & 28.94 \\
\smallskip
39 &  56.5 & 1.07 & $ 9.0_{-0.4}^{+0.4}$  & $1.53_{-0.03}^{+0.03}$ & $0.440_{-0.019}^{+0.020}$ &   25.83 & 33.50 & 1.07 & $1.50_{-0.02}^{+0.02}$ & $0.424_{-0.008}^{+0.008}$ & 26.00 & 33.28 \\
\smallskip
41 &  57.9 & 1.18 & $10.2_{-0.3}^{+0.3}$  & $1.61_{-0.02}^{+0.02}$ & $0.396_{-0.012}^{+0.012}$ &   20.21 & 27.45 & 1.35 & $1.49_{-0.01}^{+0.01}$ & $0.335_{-0.004}^{+0.004}$ & 20.79 & 26.54 \\
\smallskip
42 &  59.8 & 1.06 & $ 8.9_{-0.4}^{+0.4}$  & $1.58_{-0.03}^{+0.04}$ & $0.356_{-0.015}^{+0.016}$ &   19.30 & 25.46 & 1.07 & $1.56_{-0.02}^{+0.02}$ & $0.344_{-0.007}^{+0.007}$ & 19.41 & 25.30 \\
\smallskip
44 &  63.3 & 1.00 & $10.0_{-0.4}^{+0.4}$  & $1.62_{-0.03}^{+0.04}$ & $0.337_{-0.015}^{+0.016}$ &   17.08 & 23.19 & 1.07 & $1.51_{-0.02}^{+0.02}$ & $0.290_{-0.006}^{+0.006}$ & 17.54 & 22.51 \\
\smallskip
45 &  74.7 & 1.03 & $ 9.9_{-0.5}^{+0.5}$  & $1.61_{-0.05}^{+0.05}$ & $0.229_{-0.013}^{+0.014}$ &   11.73 & 15.86 & 1.08 & $1.51_{-0.03}^{+0.03}$ & $0.199_{-0.005}^{+0.005}$ & 12.01 & 15.43 \\
\smallskip
46 &  76.1 & 0.92 & $ 9.7_{-0.5}^{+0.5}$  & $1.62_{-0.05}^{+0.05}$ & $0.213_{-0.012}^{+0.013}$ &   10.77 & 14.58 & 0.95 & $1.54_{-0.03}^{+0.03}$ & $0.188_{-0.005}^{+0.005}$ & 11.00 & 14.22 \\
\smallskip
47 &  78.4 & 1.10 & $ 9.6_{-0.5}^{+0.6}$  & $1.63_{-0.05}^{+0.05}$ & $0.174_{-0.011}^{+0.012}$ &    8.75 & 11.84 & 1.12 & $1.55_{-0.03}^{+0.03}$ & $0.156_{-0.005}^{+0.005}$ &  8.93 & 11.59 \\
\smallskip
48 &  80.7 & 1.06 & $10.3_{-0.5}^{+0.5}$  & $1.61_{-0.05}^{+0.05}$ & $0.166_{-0.010}^{+0.011}$ &    8.46 & 11.51 & 1.14 & $1.48_{-0.03}^{+0.03}$ & $0.138_{-0.004}^{+0.004}$ &  8.75 & 11.12 \\
\smallskip
49 &  82.2 & 1.10 & $ 9.9_{-0.7}^{+0.7}$  & $1.62_{-0.06}^{+0.06}$ & $0.315_{-0.024}^{+0.026}$ &   15.94 & 21.62 & 1.13 & $1.52_{-0.03}^{+0.03}$ & $0.273_{-0.009}^{+0.009}$ & 16.34 & 21.02 \\
\smallskip
50 &  84.4 & 0.91 & $10.2_{-0.7}^{+0.7}$  & $1.58_{-0.06}^{+0.06}$ & $0.121_{-0.009}^{+0.010}$ &    6.46 &  8.67 & 0.96 & $1.46_{-0.03}^{+0.03}$ & $0.102_{-0.003}^{+0.003}$ &  6.66 &  8.42 \\
\smallskip
51 &  86.2 & 0.91 & $ 8.6_{-0.8}^{+0.8}$  & $1.50_{-0.07}^{+0.07}$ & $0.095_{-0.008}^{+0.009}$ &    5.88 &  7.52 & 0.91 & $1.49_{-0.04}^{+0.04}$ & $0.095_{-0.004}^{+0.004}$ &  5.88 &  7.51 \\
\hline\noalign{\smallskip}
\end{tabular} 
\end{center}
\end{table}

\clearpage\pagebreak

\setcounter{table}{2}
\begin{table}
\caption{Spectral parameters obtained by fitting the \swift/XRT spectra. (cont.)}
\begin{center}
\begin{tabular}{rrrlllrrrllrr}
\hline\noalign{\smallskip}
obs$^{\dagger}$&   time$^{*}$&  $\chi^2_{\mr{red}}$ & n$_H^{\#}$                      &  $\Gamma$                    &  norm$^{+}$                     & flux$_{\mr{abs}}^{\ddagger}$     & flux$_{\mr{unabs}}^{\ddagger}$ &  $\chi^{2 \spadesuit}_{\mr{red}}$ &  $\Gamma^{\spadesuit}$                    &  norm$^{+,\spadesuit}$                     & flux$_{\mr{abs}}^{\ddagger,\spadesuit}$     & flux$_{\mr{unabs}}^{\ddagger,\spadesuit}$ \\
 \hline\noalign{\smallskip}
52 &  88.2 & 0.88 & $ 8.7_{-0.6}^{+0.7}$  & $1.54_{-0.06}^{+0.06}$ & $0.092_{-0.007}^{+0.008}$ &    5.33 &  6.91 & 0.88 & $1.53_{-0.04}^{+0.04}$ & $0.090_{-0.003}^{+0.003}$ &  5.34 &  6.89 \\
\smallskip
53 &  90.6 & 0.99 & $ 8.6_{-0.8}^{+0.8}$  & $1.51_{-0.07}^{+0.07}$ & $0.081_{-0.007}^{+0.008}$ &    4.91 &  6.30 & 0.99 & $1.51_{-0.04}^{+0.04}$ & $0.081_{-0.003}^{+0.004}$ &  4.91 &  6.30 \\
\smallskip
54 &  92.5 & 0.93 & $10.1_{-0.9}^{+1.0}$  & $1.59_{-0.08}^{+0.08}$ & $0.076_{-0.007}^{+0.008}$ &    4.01 &  5.40 & 0.96 & $1.47_{-0.04}^{+0.04}$ & $0.065_{-0.003}^{+0.003}$ &  4.13 &  5.24 \\
\smallskip
55 &  94.3 & 1.03 & $ 9.0_{-0.9}^{+0.9}$  & $1.47_{-0.08}^{+0.08}$ & $0.059_{-0.006}^{+0.007}$ &    3.75 &  4.78 & 1.03 & $1.44_{-0.05}^{+0.05}$ & $0.056_{-0.003}^{+0.003}$ &  3.78 &  4.75 \\
\smallskip
57 &  98.3 & 0.94 & $ 9.5_{-2.5}^{+3.1}$  & $1.51_{-0.24}^{+0.27}$ & $0.046_{-0.012}^{+0.018}$ &    2.76 &  3.59 & 0.91 & $1.44_{-0.14}^{+0.14}$ & $0.042_{-0.006}^{+0.006}$ &  2.81 &  3.54 \\
\smallskip
58 & 100.0 & 1.09 & $10.3_{-1.2}^{+1.3}$  & $1.49_{-0.10}^{+0.11}$ & $0.045_{-0.006}^{+0.007}$ &    2.73 &  3.57 & 1.12 & $1.37_{-0.05}^{+0.05}$ & $0.038_{-0.002}^{+0.002}$ &  2.82 &  3.47 \\
\smallskip
59 & 102.4 & 1.23 & $ 9.6_{-1.2}^{+1.3}$  & $1.41_{-0.11}^{+0.11}$ & $0.035_{-0.005}^{+0.006}$ &    2.44 &  3.09 & 1.23 & $1.34_{-0.06}^{+0.06}$ & $0.032_{-0.002}^{+0.002}$ &  2.49 &  3.05 \\
\smallskip
60 & 104.0 & 1.23 & $11.9_{-1.3}^{+1.4}$  & $1.57_{-0.11}^{+0.11}$ & $0.043_{-0.006}^{+0.007}$ &    2.26 &  3.10 & 1.36 & $1.34_{-0.05}^{+0.05}$ & $0.031_{-0.002}^{+0.002}$ &  2.39 &  2.92 \\
\smallskip
61 & 106.4 & 1.10 & $ 8.7_{-1.3}^{+1.4}$  & $1.35_{-0.12}^{+0.12}$ & $0.026_{-0.004}^{+0.005}$ &    2.00 &  2.45 & 1.09 & $1.35_{-0.07}^{+0.07}$ & $0.026_{-0.002}^{+0.002}$ &  2.00 &  2.45 \\
\smallskip
62 & 108.2 & 0.94 & $ 8.8_{-1.4}^{+1.6}$  & $1.40_{-0.12}^{+0.13}$ & $0.024_{-0.004}^{+0.004}$ &    1.71 &  2.13 & 0.93 & $1.39_{-0.07}^{+0.07}$ & $0.024_{-0.002}^{+0.002}$ &  1.71 &  2.13 \\
\smallskip
63 & 110.8 & 0.87 & $11.4_{-1.5}^{+1.6}$  & $1.51_{-0.12}^{+0.13}$ & $0.028_{-0.004}^{+0.005}$ &    1.66 &  2.21 & 0.95 & $1.32_{-0.06}^{+0.06}$ & $0.021_{-0.001}^{+0.001}$ &  1.74 &  2.12 \\
\smallskip
65 & 114.7 & 1.37 & $13.6_{-3.7}^{+4.3}$  & $1.70_{-0.29}^{+0.31}$ & $0.037_{-0.012}^{+0.018}$ &    1.58 &  2.32 & 1.47 & $1.35_{-0.13}^{+0.13}$ & $0.023_{-0.003}^{+0.003}$ &  1.73 &  2.12 \\
\hline\noalign{\smallskip}
\end{tabular} 
\end{center}
\normalsize
Notes: $^{*}$: days since June 10th 2015 00:00:00.000 UTC\\
$^{\dagger}$: \swift\ obs id: 000338110xx\\
$^{\ddagger}$: $10^{-10}$ erg cm$^{-2}$ s$^{-1} $\\
$^{+}$: photons keV$^{-1}$ cm$^{-2}$ s$^{-1} $   at 1 keV\\
$^{\#}$: $10^{21}$ cm$^{-2}$\\
$^{\spadesuit}$: n$_H$ fixed at the average of $8.6\times10^{21}$ cm$^{-2}$
\end{table}

\begin{table}
\caption{\label{Tab:sw_qpo}Parameters of the PDS of \swift\ observations that show type-C QPOs}
\begin{center}
\begin{tabular}{rllllllll}
\hline\noalign{\smallskip}
 & \multicolumn{2}{c}{BLN} & \multicolumn{3}{c}{QPO} & \multicolumn{3}{c}{harmonic/ add.~comp.} \\
\hline\noalign{\smallskip}
 obs$^{\dagger}$ & $\Delta$ [Hz]& rms [\%]& $\nu$ [Hz]& $\Delta$ [Hz]& rms [\%]& $\nu$ [Hz]& $\Delta$ [Hz]& rms [\%] \\
\hline\noalign{\smallskip}
37 &  $0.702_{-0.124}^{+0.167}$  &   $18.7_{-1.5}^{+1.3}$  &   $0.183_{-0.004}^{+0.005}$  &   $0.007_{-0.003}^{+0.004}$  &   $ 8.9_{-1.5}^{+1.4}$  &   $0                      $  &   $0.028_{-0.010}^{+0.012}$  &   $9.7_{-1.7}^{+1.4}$ \\
\smallskip
38 &  $0.154_{-0.041}^{+0.044}$  &   $18.3_{-2.3}^{+2.4}$  &   $0.182_{-0.012}^{+0.009}$  &   $0.015_{-0.013}^{+0.015}$  &   $ 9.1_{-2.5}^{+2.1}$  &   $0.420_{-0.019}^{+0.020}$  &   $0.039_{-0.018}^{+0.026}$  &   $8.2_{-2.0}^{+1.8}$ \\
\smallskip
39 &  $0.258_{-0.046}^{+0.053}$  &   $21.5_{-1.5}^{+1.3}$  &   $0.215_{-0.013}^{+0.010}$  &   $0.018_{-0.007}^{+0.008}$  &   $10.3_{-1.9}^{+1.8}$  &   $                       $  &   $                       $  &   $$\\
\smallskip
41 &  $0.325_{-0.034}^{+0.042}$  &   $23.3\pm0.9        $  &   $0.192_{-0.006}^{+0.005}$  &   $0.021_{-0.006}^{+0.007}$  &   $11.2_{-1.4}^{+1.3}$  &   $0.399_{-0.001}^{+0.002}$  &   $<0.004                 $  &   $4.4_{-0.9}^{+0.7}$ \\
\smallskip
42 &  $0.454_{-0.098}^{+0.175}$  &   $19.1_{-1.8}^{+1.5}$  &   $0.187\pm0.010          $  &   $0.020_{-0.007}^{+0.009}$  &   $11.9\pm2.0        $  &   $                       $  &   $                       $  &   $$\\
\smallskip
44 & $0.926_{-0.239}^{+0.262}$  &   $19.1_{-1.6}^{+1.7}$  &   $0.194\pm0.007          $  &   $0.018_{-0.008}^{+0.007}$  &   $11.2_{-1.5}^{+1.2}$  &   $0.087_{-0.005}^{+0.004}     $  &   $0.008_{-0.004}^{+0.005}$  &   $7.6_{-1.4}^{+1.2}$  \\
\smallskip
47 &  $3.186_{-1.787}^{+3.139}$  &   $29.1_{-4.9}^{+4.4}$  &   $0.176_{-0.005}^{+0.006}$  &   $0.010_{-0.010}^{+0.021}$  &   $10.2_{-4.2}^{+2.4}$  &   $0.070\pm0.030          $  &   $0.040_{-0.025}^{+0.181}$  &   $>9.5$ \\
\hline\noalign{\smallskip} 
\end{tabular} 
\end{center}
Notes: \\
$^{\dagger}$: \swift\ obs id: 000338110xx
\end{table}

\bsp

\label{lastpage}

\end{document}